# Phonon-roton modes of liquid $^4$He beyond the roton in MCM-41

R.T. Azuah,[1,2] S.O. Diallo,[3] M. A. Adams,[4] O. Kirichek,[4] and H. R. Glyde[5]

[1]*NIST Center for Neutron Research, Gaithersburg, Maryland 20899-8562, USA*
[2]*Department of Materials Science and Engineering,
University of Maryland, College Park, Maryland 20742-2115, USA*
[3]*Quantum Condensed Matter Division, Oak Ridge National Laboratory, Oak Ridge, TN 37831, USA*
[4]*ISIS Spallation Neutron Source, Rutherford Appleton Laboratory, Chilton, Didcot, OX11 0QX, United Kingdom*
[5]*Department of Physics and Astronomy, University of Delaware, Newark, Delaware 19716-2593, USA*
(Dated: May 18, 2013)

We present neutron scattering measurements of the phonon-roton (P-R) mode of superfluid $^4$He confined in 47 Å MCM-41 at $T = 0.5$ K at wave vectors, $Q$, beyond the roton wave vector ($Q_R = 1.92$ Å$^{-1}$). Measurements beyond the roton require access to high wave vectors (up to $Q = 4$ Å$^{-1}$) with excellent energy resolution and high statistical precision. The present results show for the first time that at $T = 0.5$ K the P-R mode in MCM-41 extends out to wave-vector $Q \simeq 3.6$ Å$^{-1}$ with the same energy and zero width (within precision) as observed in bulk superfluid $^4$He. Layer modes in the roton region are also observed. Specifically, the P-R mode energy, $\omega_Q$, increases with $Q$ for $Q > Q_R$ and reaches a plateau at a maximum energy $\omega_Q = 2\Delta$ where $\Delta$ is the roton energy, $\Delta = 0.74 \pm 0.01$ meV in MCM-41. This upper limit means the P-R mode decays to two rotons when its energy exceeds $2\Delta$. It also means that the P-R mode does not decay to two layers modes. If the P-R could decay to two layer modes, $\omega_Q$ would plateau at a lower energy, $\omega_Q = 2\Delta_L$ where $\Delta_L = 0.60$ meV is the energy of the roton like minimum of the layer mode. The observation of the P-R mode with energy up to $2\Delta$ shows that the P-R mode and the layer modes are independent modes with apparently little interaction between them.

Bulk superfluid $^4$He supports a well-defined phonon-roton (P-R) mode at wave vectors, $Q$, from $Q = 0$ up to $Q = 3.6$ Å$^{-1}$. The P-R mode may be said to be a mode in the density response of a Bose liquid.[1–3] At low temperatures, below 0.7 K, the mode is so sharply defined that its width is too small to be observed. When there is Bose-Einstein condensation (BEC), as in the superfluid phase, the mode is sharply defined because there are no low lying modes to which the mode can decay. Notably, when there is BEC, the single particle mode does not lie at low energy. Rather the single particle response has the same phonon-like poles (energies) as the density response so that there is only a single mode in the fluid.[4–6] The P-R mode is therefore the only mode and can decay only to other P-R modes. At low temperatures the single mode therefore remains sharply defined until its energy is high enough that it can decay to two other modes. Specifically, if the P-R mode energy exceeds $2\Delta$, where $\Delta$ is the P-R mode energy at the roton minimum, the mode has sufficient energy to decay spontaneously to two rotons[7] and the mode ceases to be well-defined. The P-R mode energy dispersion curve in bulk helium (and in helium confined in MCM-41 as observed here) is shown in Fig. 1 with the energies $\Delta$ and $2\Delta$ identified.

The goal of the present work is to explore whether this basic picture remains true for liquid $^4$He confined in nanoporous media. Does the P-R mode continue to exist out to $Q = 3.6$ Å$^{-1}$ when the liquid is confined to nanoscales and is in a disordered environment? Is the P-R mode energy and line width at wave vectors beyond the roton ($1.92 < Q < 3.6$ Å$^{-1}$) modified by disorder? Specifically, at $Q = 1.92$ Å$^{-1}$ the full width at half maximum (FWHM) of the roton mode in bulk superfluid investigated here, $T = 0.5$ K, the FWHM at the roton in bulk liquid $^4$He extrapolates to $10^{-2}$ $\mu$eV, undetectably small. Does the P-R mode width in MCM-41 at $T = 0.5$ K remain undetectably small at wave vectors beyond the roton? Particularly, does the P-R mode energy reach $2\Delta$ as in the bulk or can the P-R mode decay to two layer modes which exist in porous media and which have a lower energy? In the present MCM-41 the layer mode has a roton like minimum energy of $\Delta_L = 0.60$ meV ($2\Delta_L = 1.20$ meV) rather than $2\Delta = 1.48$ meV for two rotons. If the P-R mode can decay to two layer modes, we expect the P-R mode to broaden when its energy exceeds $2\Delta_L$ with perhaps an energy upper limit of $2\Delta_L$.

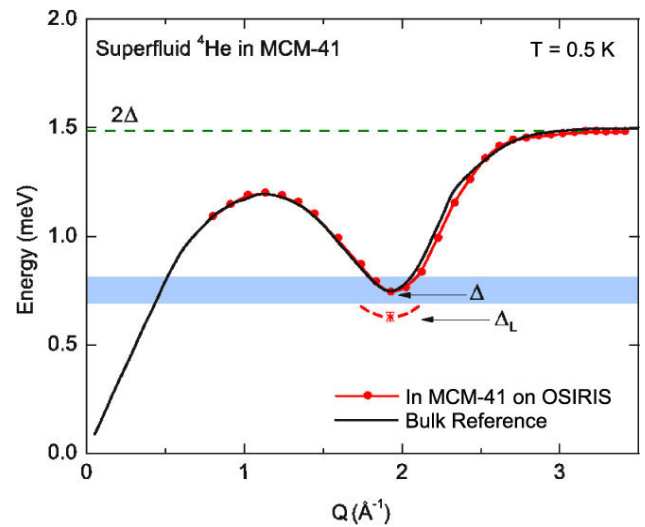

FIG. 1: The phonon-roton mode energy dispersion curve of (a) bulk superfluid $^4$He at low temperature (black solid line) from Ref. 8 ($0 < Q < 2.4$ Å$^{-1}$) and Ref. 9 ($2.4$ Å$^{-1} < Q < 3.6$ Å$^{-1}$) and (b) superfluid $^4$He in fully filled MCM-41 at $T = 0.5$ K observed in the present measurements on OSIRIS (solid red points and red line), both at saturated vapor pressure (SVP). Also shown is the layer mode energy at roton wave vectors ($Q \simeq 1.95$ Å$^{-1}$) in MCM-41 (red asterisk and dashed red line) at SVP. $\Delta$ and $\Delta_L$ denote the roton and layer roton energies, respectively. At higher $Q$, the P-R mode energies, those of Ref. 9 and the present, plateau at and do not exceed $2\Delta$. The blue band denotes the energy range in which the roton will be observed at other wave vectors because of multiple scattering from the roton and the MCM-41.

$^4$He is about,[10–12] 1 $\mu$eV at 1 K. At the temperatures

The dynamical structure factor (DSF), $S(Q,\omega)$, of liquid $^4$He at saturated vapor pressure (SVP) at lower wave vectors, $Q \leq 2.4$ Å$^{-1}$ has been investigated in several porous media; aerogel,[13–20] Xerogel,[21,22] Vycor[20,23–25] gelsil,[26–31] of 25, 34 and 44 Å mean pore diameter (MPD), MCM-41[25,30–33] of 35 Å and 47 Å MPD and FSM-28.[34] When the porous media is fully filled, the P-R mode energy in all porous media investigated to date is the same as in bulk superfluid $^4$He. This is both at low temperature and as a function of temperature within precision[35]. The P-R mode energy is also the same as in bulk helium as a function of pressure within observable precision.[28,30,31,33] At low temperature and at SVP the mode width is unobservably small, as in the bulk liquid. At a significantly higher pressure (34 bar) a small mode width at $T = 0.2$ K in MCM-41 (47 Å diameter) has been observed.[31] At partial filling of porous media[29,35] and in helium films on surfaces[36,37], the P-R mode energies differ from the bulk values. At partial filling the P-R mode energy is lower at wave vectors $Q \simeq 1$ Å$^{-1}$ (the maxon region) and higher in the roton region[19,38] as if the films have a lower density than the bulk density. Rotons in bulk helium have been recently observed using Microwaves.[39]

In porous media[20,22–25,27,29–32,35] and in helium films,[36,40,41] a layer mode is also observed. This is a mode that propagates in the 1-4 liquid layers adjacent to the porous media walls. These liquid layers lie on top of the solid helium layers (1-2 layers) that coat the pore walls. The layer mode is observable chiefly at wave vectors in the roton region (1.7 < Q < 2.2 Å$^{-1}$). The layer mode energy has a roton like minimum at $Q \simeq 1.9 - 2.1$ Å$^{-1}$. The minimum energy is typically $\Delta_L \simeq 0.5$-0.65 meV which is significantly lower than the P-R mode roton minimum, $\Delta = 0.742 \pm 0.001$ eV in bulk liquid $^4$He. Both $\Delta$ and $\Delta_L$ decrease with increasing pressure[31]. In this context, our goal is to extend measurements of $S(Q,\omega)$ in porous media to wave vectors beyond the roton (1.92 < Q < 3.6 Å$^{-1}$) and to determine whether the P-R mode can decay to two layer modes or not.

In section 2, we describe the MCM-41 sample, the sample cell and the OSIRIS neutron scattering instrument used to measure $S(Q,\omega)$. We present the measurement results for $S(Q,\omega)$ and the P-R mode energy, $\omega_Q$, FWHM and intensity, $Z_Q$ in section 3. The results are discussed in section 4 and summarized in section 5.

## I. EXPERIMENT

### A. Porous Media Samples

The present MCM-41 sample is the same as used and described previously by us[25,42,43]. It was synthesized at the Laboratoire de Matériaux Minéraux, UMR-CNRS, Mulhouse, France following the procedure of Cormka et al.[44]. The sample is a white powder of micrometer grain size. The silica grains contain parallel cylindrical pores ordered in a hexagonal lattice. N$_2$ isotherms, performed and analyzed by Mulhouse using the standard Barrett, Joyner and Halenda (BJH) model,[45] yielded a mean pore diameter of 47 Å with a narrow pore diameter distribution of half width at half maximum (HWHM) of 1.5 Å. Analysis using the Brunauer, Emmett and Teller (BET) model[46] indicated a pore volume of $v_P = 0.931$ cm$^3$/g. From diffraction measurements a lattice constant $a = 63$ Å for the hexagonal lattice formed by the pores was found. Because of the presence of silanol and of incomplete polycondensation of the silica, the silica density was $\rho = 1.8$ g/cm$^3$ rather than the usual 2.1 g/cm$^3$. Thus the volume of the silica is $v_{Sil} = 1/\rho = 0.55$ cm$^3$/g giving a sample porosity of $p = v_P/(v_P + v_{Sil}) = 63$ %.

$^4$He isotherms performed by us (see Fig. 1 of Ref. 25) show that 47 Å MCM-41 pores are fully filled when 40 mmol at STP/g of $^4$He is added to the powder. At this filling the vapor pressure above the $^4$He reaches the bulk saturated vapor pressure (SVP) value indicating that bulk liquid helium is being added. Neutron scattering measurements were made at (1) one half filling, (2) full filling and (3) overfilling (70.4 mmol/gm). $^4$He isotherm measurements show that at half filling the vapor pressure of the $^4$He is very small, about 2 % of the bulk SVP for temperatures up to $T = 2.5$ K. This indicates that at half filling the $^4$He is still tightly bound (as solid and liquid layers) to the pore walls. Capillary filling, indicating filling with less tightly bound liquid begins at about 25 mmol/gm. We used filling (1) as a background measurement.

### B. Neutron Scattering Measurements

The MCM-41 sample of total mass $M_S = 15.45$ g was placed in a cylindrical aluminum cell and cooled to 0.5 K and 1.5 K in a $^3$He cryostat. The neutron scattering measurements were performed on the OSIRIS spectrometer at the ISIS facility, Rutherford Appleton Laboratory, UK. OSIRIS has an energy resolution of 100 $\mu$eV in the $Q$ range 1.5 - 4 Å$^{-1}$ investigated here. This high energy resolution is obtained at the cost of rather broad $Q$ resolution, $\Delta Q$. This trade off is excellent for measuring the energy accurately near the end of the P-R dispersion curve, $2.8 < Q < 3.6$ Å$^{-1}$ where the P-R energy, $\omega_Q$, is changing little with Q. However, in $Q$ ranges where $\omega_Q$ is changing rapidly with $Q$, the broad $Q$ resolution will lead to an apparent energy broadening since a range of $Q$ is being collected. Under these conditions, Crevecoeur et al.[47] show the total observed peak width is $W^2 = (\Delta\omega)^2 + (d\omega_Q/dQ)^2 (\Delta Q)^2$. If in addition the intensity, $Z_Q$, in the P-R peak is changing rapidly with $Q$, the peak position can be shifted away from its true value. In the wave vector range $2.2 < Q < 2.6$ Å$^{-1}$ where both $(d\omega_Q/dQ)$ and $(dZ_Q/dQ)$ are large (and $(dZ_Q/dQ)$ is negative) we expect the peak position observed on OSIRIS to be shifted below its actual value. To monitor this effect we also measured the P-R energy dispersion curve of bulk liquid $^4$He, where the energy is well known. We see indeed that the P-R energy, $\omega_Q$, of bulk liquid $^4$He observed on OSIRIS in the range $2.2 < Q < 2.6$ Å$^{-1}$ lies below the most accurate values[8,48] (see Fig. 3). We also measured the FWHM of P-R mode peak in bulk liquid at $T = 0.5$ K (where the intrinsic mode width is less than $10^{-2}$ $\mu$eV) to calibrate the effective width $W$ of OSIRIS (see Fig. 6). At the wave vectors $1.4 < Q < 1.8$ Å$^{-1}$ and $2.2 < Q < 2.6$ Å$^{-1}$ where $(d\omega_Q/dQ)$ is large, we see that the effective $W$ is large. At $Q = 1.92$ Å$^{-1}$ and $Q > 2.8$ Å$^{-1}$ where $(d\omega_Q/dQ) = 0$, the effective width $W$ reduces to the intrinsic energy resolution width of OSIRIS, 0.1 meV.

## II. RESULTS

Measurements of the inelastic scattering intensity from the sample at three fillings of the MCM-41 with liquid $^4$He at $T = 0.5$ K were made. The fillings were (1) half full pores (20 mmol/g), (2) full pores (40 mmol/g) and (3) full pores plus 30.4 mmol/g of bulk liquid between the grains of the MCM-41 powder. The intensity (1), which includes the scattering from the solid layers on the pore walls, was used as a background. We present specifically the net scattering, (2) minus (1), from liquid $^4$He lying from half to full filling to represent fully filled MCM-41 and (3) minus (2), overfilled MCM-41 to represent bulk liquid helium as a reference. The data were analysed (modes fitted to the data) at constant $\phi$ and the dispersion curves were converted to constant $Q$. Some intensities were converted to constant $Q$ for display of $S(Q,\omega)$.

Fig. 2 shows the net dynamic structure factor, $S(Q,\omega)$, of liquid $^4$He in MCM-41 at $T = 0.5$ K at three wave vectors $Q$. The top frame, $Q = 1.92$ Å$^{-1}$, is at the roton minimum. The open circles and solid black line are the data and the fit to the data, respectively. The sharply peaked solid red line is the P-R mode component (the roton at $Q = 1.92$ Å$^{-1}$). The broad solid red line centered at $\Delta_L = 0.60$ meV is the layer mode. The layer mode intensity is small because the layer modes propagate in the liquid layers adjacent to the pore walls and the net $S(Q,\omega)$ is from liquid predominantly in the center of the pores, from half full to full pores. The dashed blue line represents the scattering which creates two or more P-R modes which is small at the roton $Q$.

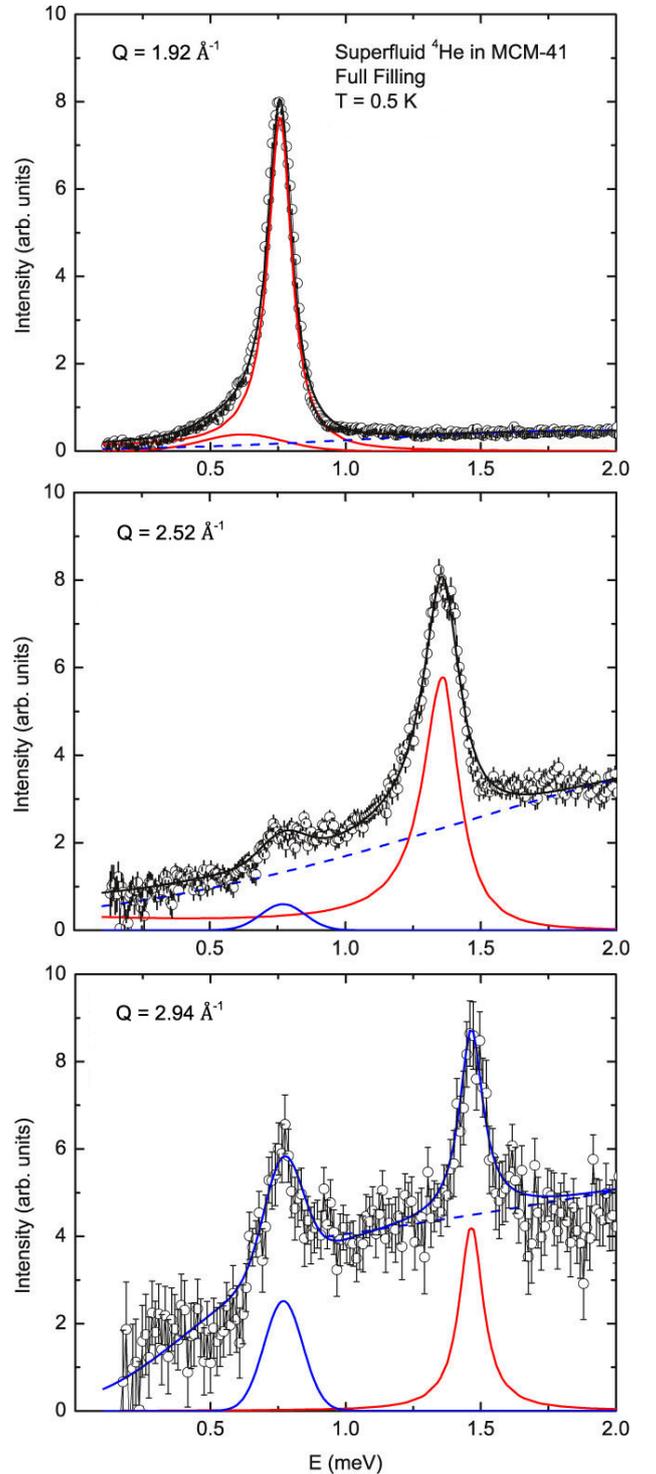

FIG. 2: The net scattering intensity, proportional to $S(Q,\omega)$, from liquid $^4$He in MCM-41 at $T = 0.05$ K and $Q = 1.92$ Å$^{-1}$, $Q = 2.52$ Å$^{-1}$, and $Q = 2.94$ Å$^{-1}$ (from top to bottom). Shown is the net $S(Q,\omega)$ from the liquid $^4$He lying between one-half filling (20 mmol/g) and full filling of the MCM-41 (40 mmol/g). The points and solid black line show the total observed intensity and the fit to the intensity, respectively. The peaked solid red line identifies the phonon-roton mode intensity. The weak, broad red line at $Q = 1.92$ Å identifies the layer mode. At $Q = 2.52$ Å$^{-1}$ and $Q = 2.94$ Å$^{-1}$ the solid blue line is the roton mode ($Q = 1.92$ Å$^{-1}$) observed at other $Q$ values as a result of multiple scattering. The dashed blue line is the multi-mode/ interacting single particle scattering background which increases rapidly with increasing $Q$.

In the middle frame of Fig. 2, $Q = 2.52$ Å$^{-1}$, the solid red line is again the P-R mode. The dashed blue line is the scattering that creates several P-R modes which increases as $Q$ increases. The solid blue line is the roton mode (normally observed at $Q = 1.92$ Å$^{-1}$) which is observable at $Q = 2.52$ Å$^{-1}$ because of multiple scattering. The multiple scattering is an inelastic scattering which creates a roton in the liquid followed (or preceded) by an elastic scattering from the MCM-41 which changes the $Q$ value. The elastic scattering from the MCM-41 means that the $Q$ selection is lost and intense modes such as the roton can appear at other $Q$ values.

In the bottom frame of Fig. 2, $Q = 2.94$ Å$^{-1}$, the intensity in the P-R mode is significantly reduced and the intensity in the "background" multimode component has increased significantly The multi-mode component can be equivalently described as scattering that creates dressed single particle excitations. At $Q = 2.94$ Å$^{-1}$ the intensity in the P-R mode is already small and P-R mode energy has reached its maximum value $2\Delta$ ($2\Delta = 1.484 \pm 0.002$ meV in bulk superfluid $^4$He). As $Q$ increases further, the intensity in the P-R mode decreases. The P-R mode in MCM-41 was last distinguishable from the dressed single particle scattering at $Q = 3.5$ Å$^{-1}$.

To extract the energy, $\omega_Q$, the FWHM, $2\Gamma_Q$, and the intensity, $Z_Q$, of the P-R mode at wave vector $Q$, we represented the net observed intensity as the sum of three parts;

$$S(Q,\omega) = S_{PR}(Q,\omega) + S_{MS}(Q,\omega) + S_B(Q,\omega). \quad (1)$$

In Eq. (1), $S_{PR}(Q,\omega)$ is the P-R mode component. It is represented by the standard damped harmonic oscillator (DHO) function written as a sum of two Lorentzians,

$$S_{PR}(Q,\omega) = \frac{Z_Q/\pi}{1 - \exp(-\hbar\omega/k_B T)} \times \left[ \frac{\Gamma_Q}{(\omega - \omega_Q)^2 + \Gamma_Q^2} - \frac{\Gamma_Q}{(\omega + \omega_Q)^2 + \Gamma_Q^2} \right]. \quad (2)$$

In Eq. (2) the $\omega_Q$, $2\Gamma_Q$ and $Z_Q$ are free parameters adjusted to give the best fit to the data. The $S_{MS}(Q,\omega)$ is the multiple scattering component. Since the scattering from the P-R mode is most intense for the roton, the multiple scattering essentially enables the strong roton intensity to appear at all Q values. Multiple scattering peaks are shown in the two bottom frames of Fig. 2. $S_{MS}(Q,\omega)$ is similarly represented by a DHO function. The $S_B(Q,\omega)$ is an adjustable sloping background contribution.

For the special case of the roton wave vector $Q = 1.92$ Å$^{-1}$, we used the model,

$$S(Q,\omega) = S_{PR}(Q,\omega) + S_L(Q,\omega) + S_B(Q,\omega), \quad (3)$$

in which $S_{PR}(Q,\omega)$ represents the roton mode, $S_L(Q,\omega)$ represents the layer mode and $S_B(Q,\omega)$ is again the background. The $S_L(Q,\omega)$ was written as a Gaussian function essentially because the layer mode is broad and approximately symmetric around its maximum value.

Because of the multiple scattering in the present data, the layer mode can be clearly separated from the P-R mode at the roton wave vector only. Specifically, the minimum energy of the layer mode is $\Delta_L = 0.6$ meV. This minimum occurs approximately[32] at the roton wave vector $Q = 1.92$ Å$^{-1}$. Thus at $Q \simeq 1.9$ Å$^{-1}$ the layer mode lies at an energy well below the P-R mode (below the roton energy $\Delta \simeq 0.74$ meV) and the layer mode can be distinguished from the roton mode at this $Q$ value. The layer mode energy is higher at other $Q$ values[32]. At other $Q$ values we still have intensity at the roton energy arising from multiple scattering from the roton, shown as a blue band in Fig. 1. Essentially at other $Q$ values we cannot separate the weak layer mode intensity from the multiple scattering from the roton. Thus we attempted to fit the layer mode and obtain the layer mode energy at the roton $Q = 1.92$ Å$^{-1}$ only.

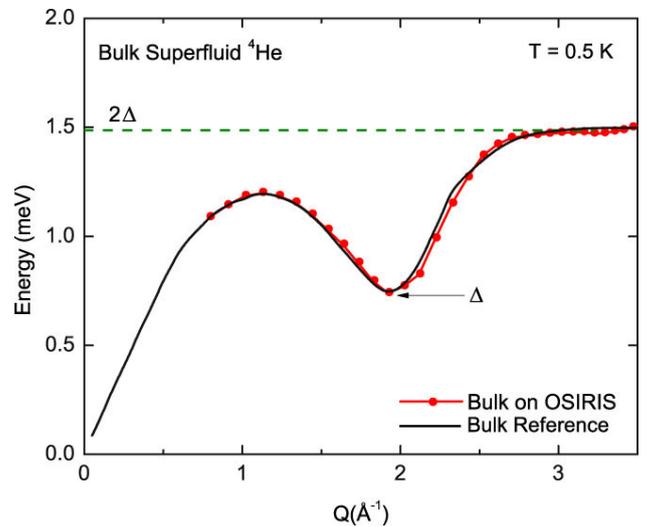

FIG. 3: The phonon-roton mode energy dispersion curve of bulk superfluid $^4$He at SVP: (1) reference values from Ref. 8 ($0 < Q < 2.4$ Å$^{-1}$) and Ref. 9 ($2.4$ Å$^{-1} < Q < 3.6$ Å$^{-1}$) (black solid line) and (2) as observed in the present measurements on OSIRIS (solid red points and red line).

Fig. 3 compares the present values of the P-R mode energy of bulk liquid $^4$He, $\omega_Q$, observed on OSIRIS with reference values.[8,48] The aim is to calibrate OSIRIS. The agreement is good, within the error bars given by the size of the points, except in the wave vector range $2.1 < Q < 2.6$ Å$^{-1}$. In this range the intensity, $Z_Q$, in the P-R mode is decreasing rapidly with increasing $Q$. This change, combined with the rather broad $Q$ resolution of OSIRIS, shifts the apparent peak position observed on OSIRIS to lower energies, as discussed in section IIB. The origin of this small discrepancy is therefore understood.

Fig. 1 shows the present values of the P-R mode energy in liquid $^4$He in MCM-41 compared with reference values[8,48] in the bulk. By comparing Figs. 1 and 3, we see that the P-R mode energy of liquid $^4$He in MCM-41 and in the bulk are the same within error ($\pm$ 0.02 meV). At lower $Q$ values, $Q < 2.4$ Å$^{-1}$, the $\omega_Q$ of liquid $^4$He in fully filled porous media and in the bulk have been found to the same[20,21,24,27,29,32,35]. The present data extends this finding out to the end point of the P-R dispersion curve, $Q = 3.6$ Å$^{-1}$. The layer mode energy, $\Delta_L = 0.60 \pm 0.03$ meV, observed here at the roton wave vector also agrees with previously observed values.[32]

Fig. 4 shows the apparent FWHM, $2\Gamma_Q$, of the P-R mode in bulk liquid $^4$He at $T = 0.5$ K as observed on OSIRIS. Since the intrinsic width of the mode at 0.5 K is less than $10^{-4}$ meV, the observed width represents the resolution width of OSIRIS. As discussed in section IIB, this resolution width is a function of $Q$ in liquid $^4$He and has minima of 0.10 meV at $Q$ values where $(d\omega_Q/dQ) \simeq 0$ (e.g. at $Q > 2.8$ Å$^{-1}$).

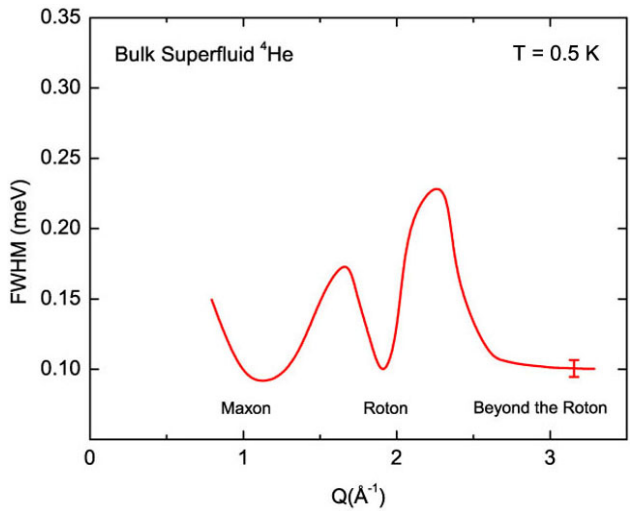

FIG. 4: The full width at half maximum (FWHM) of the phonon-roton mode of bulk superfluid $^4$He at $T = 0.5$ K observed on OSIRIS. The FWHM shown represents the OSIRIS instrument resolution width since the intrinsic P-R mode FWHM at 0.5 K is[10,11] less than $10^{-4}$ meV. The peaks in the FWHM at $Q \simeq 1.7$ Å and $Q \simeq 1.92$ Å arise from the rather broad resolution in $Q$ and the rapidly changing P-R energy with $Q$ at these $Q$ values.

The top frame of Fig. 5 compares the FWHM of liquid $^4$He confined in MCM-41 with that in the bulk liquid both at 0.5 K, both observed on OSIRIS. There we see that the width is the same within the error bar. Thus confining liquid $^4$He in MCM-41 does not open up any new decay channels for the P-R mode even at larger wave vector. Particularly, the P-R mode does not decay to layer modes. If it did, we would expect a significant broadening of the P-R mode beginning at $Q \simeq 2.5$ Å$^{-1}$ where the P-R mode energy exceeeds $2\Delta_L = 1.20$ meV, the energy needed to decay spontaneously to two layer modes. The bottom frame of Fig. 5 shows the FWHM of the liquid in MCM-41 at $T = 1.5$ K. There we see a clear thermal broadening of the P-R mode at $T = 1.5$ K. We also see the mode wave vector (e.g. the roton wave vector $Q_R$) has shifted to higher $Q$ values at higher temperature, as is observed in the bulk.[49,50]

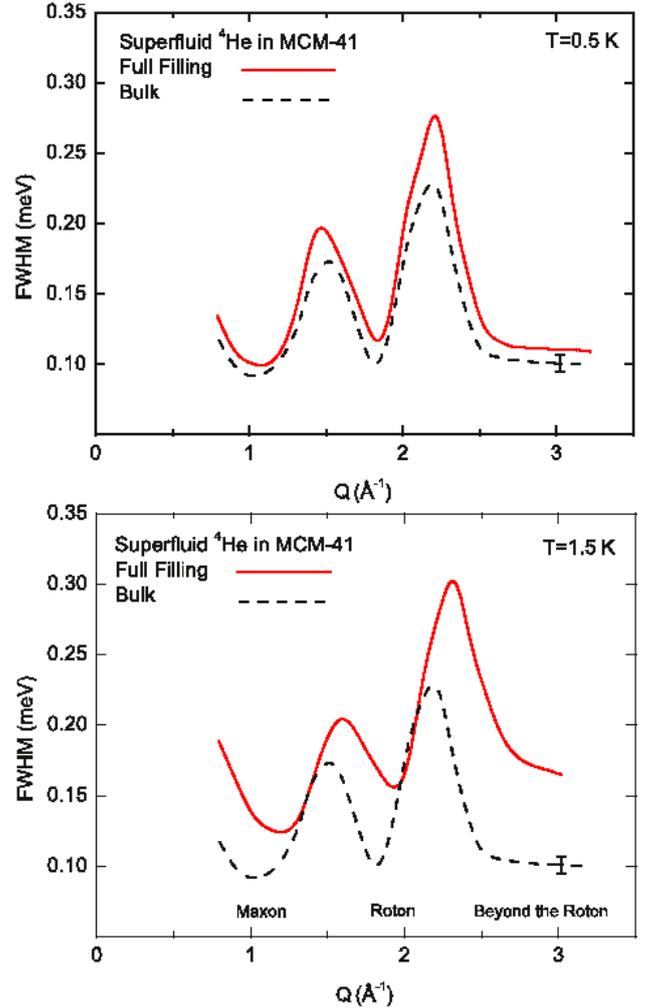

FIG. 5: Top: The FWHM of superfluid $^4$He at $T = 0.5$ K in MCM-41 at full filling (solid red line) compared with that in bulk superfluid $^4$He (dashed black line), the instrument resolution. The difference lies within the error bar. Bottom: The FWHM of superfluid $^4$He in MCM-41 (solid red line) at $T = 1.5$ K compared with that in bulk superfluid $^4$He at $T = 0.5$ K. A thermal broadening of the mode and a shift in the roton $Q$ between 0.5 K and 1.5 K is observed.

Fig. 6 compares the intensity, $Z_Q$, in the P-R mode in the bulk observed on OSIRIS with references values.[51] The magnitude of the intensity observed on OSIRIS was adjusted so that the magnitude of the two intensities was the same at their peak. With this adjustment the two intensities agree well, notably their $Q$ dependence. The measurements on OSIRIS provide new accurate values of the intensity for bulk liquid $^4$He for $Q > 2.4$ Å$^{-1}$. Fig. 7 compares the intensity, $Z_Q$, in the P-R mode of liquid $^4$He in MCM-41 with that in the bulk, both observed on OSIRIS. The magnitude of $Z_Q$ in MCM-41 was adjusted so that the two intensities had the same peak height. The $Z_Q$ in MCM-41 is clearly the same as in the bulk except at higher $Q > 2.3$ Å$^{-1}$ where the intensity in the P-R mode in MCM-41 may be somewhat smaller. As seen in Fig. 7, the intensity in the P-R mode is small at higher $Q$. A small error in the large background intensity at higher $Q$ in MCM-41, as seen in Fig. 2, could account for this difference.

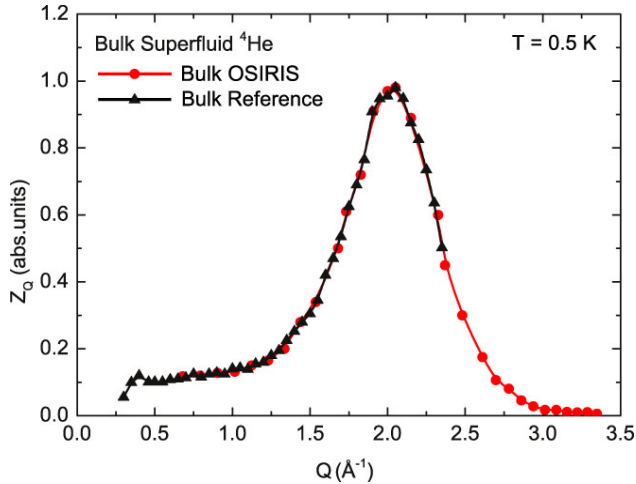

FIG. 6: The intensity, $Z_Q$, versus $Q$ of the phonon-roton mode of bulk superfluid $^4$He: (1) reference values as observed[51] on IN6 at Institut Laue Langevin (black triangles and black line) and (2) observed in the present measurements on OSIRIS (solid red points and red line). The intensity scale of the present $Z_Q$ was adjusted so that the peak height of the two intensities coincide. The two $Z_Q$ clearly agree well.

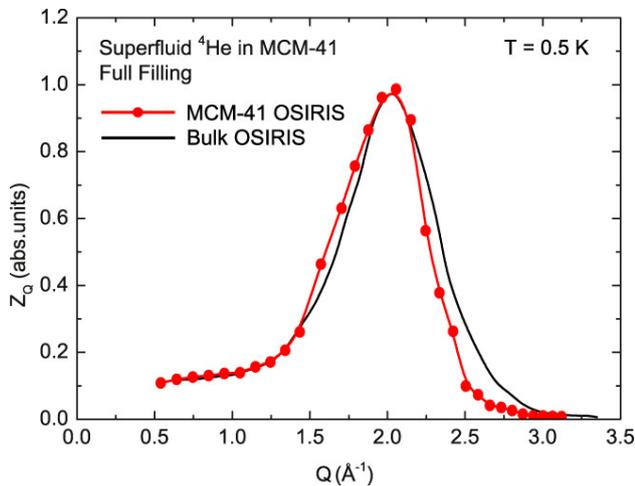

FIG. 7: The intensity, $Z_Q$ vs $Q$ observed in the present measurements on OSIRIS of the phonon-roton mode of superfluid $^4$He in (1) fully filled MCM-41 (solid red points and red line) and (2) in bulk superfluid $^4$He (black line) from Fig. 4. The intensity of $Z_Q$ in MCM-41 was adjusted so that the peak height of the two intensities coincide.

## III. DISCUSSION

In the wave vector range $2.5 < Q < 5$ Å$^{-1}$, the DSF of liquid $^4$He may be written in form,[9,52,53]

$$S(Q,\omega) = S_1(Q,\omega) + \sum_i S_i(Q,\omega) + S_R(Q,\omega), \quad (4)$$

where $S_1(Q,\omega)$ represents scattering processes that create a single P-R mode. The $S_i(Q,\omega)$ represent processes that create $i = 2$ or more P-R modes. There are in addition presumably interference terms between the single and multimode components of $S(Q,\omega)$ as have been clearly identified in the solid helium, for example. The $S_R(Q,\omega)$ is the regular component which does not involve any modes or the condensate and is largely independent of temperature. The observed $S(Q,\omega)$ of bulk liquid at SVP[9] and at 20 bars[52] shows these three components in Eq.(4) clearly, as do models[53] of $S(Q,\omega)$. At higher $Q$ values ($Q > 5$ Å$^{-1}$) a description based on scattering that excites single atoms (dressed single atoms), that leads into the Impulse Approximation at very high $Q$, is expected to be a better representation.

$S_1(Q,\omega)$ consists of a sharp single mode component $S_{PR}(Q,\omega)$ that lies at energies $\omega < 2\Delta$ plus a broad component that lies at energies $\omega > 2\Delta$ in the multi-phonon band. The P-R mode is defined as the sharp component, $S_{PR}(Q,\omega)$. The broad component of $S_1(Q,\omega)$ arises from the spontaneous decay of the single mode into two or more P-R modes which becomes possible for energy $\omega > 2\Delta$. As $Q$ increases beyond $Q \simeq 2$ Å$^{-1}$, the intensity in the sharp component $S_{PR}(Q,\omega)$ decreases and the intensity is transferred to the broad component until the intensity in the sharp component goes to zero at $Q = 3.6$ Å$^{-1}$ which defines the end point of the P-R mode. The intensity in the sharp component, $Z_Q$, is depicted here in bulk and in confined liquid $^4$He in Fig. 7. At $Q \geq 3.6$ Å$^{-1}$ the intensity in $S_1(Q,\omega)$ is entirely in the broad component at $\omega > 2\Delta$.

In the present measurements we have focused entirely on the P-R mode, the sharp component of $S_1(Q,\omega)$, for $2 < Q < 4$ Å$^{-1}$. When liquid $^4$He is confined in MCM-41, the energy, $\omega_Q$, and intensity, $Z_Q$, in the mode at $T = 0.5$ K is indistinguishable from the bulk liquid values. The width of the P-R mode is too small to be observed (FWHM $\leq 10$ $\mu$eV). The small width shows that confinement to nanoscales and disorder opens no new decay processes. The P-R mode apparently can decay only to other P-R modes as in the bulk. Surprisingly, the P-R mode does not decay to the layer modes that propagate in the liquid layers adjacent to the porous media walls, at least within present precision. This means that the P-R and layer modes are quite independent modes and interact only weakly. This is the case even though the P-R mode and the layer mode both propagate in the 2-5 liquid layers adjacent to the media walls (i.e. in the same region of the liquid). For example, in liquid $^3$He, the density mode is very broad because it can decay to particle-hole excitations. Similarly, possible variations in the liquid density within the pores or near walls do not broaden the modes. Nor is finite size apparently important, presumably because the mode wavelength $\lambda = 2\pi/Q \leq 3$ Å is short compared to the confinement lengths (20 Å). For all these reasons, it is surprising that the width of the P-R mode in MCM-41 at higher $Q$ values is undetectably small.

The intensity in the layer mode at $Q = 1.92$ Å shown in the top frame of Fig. 2 is significantly smaller than that observed in previous measurements in comparable small pore porous media.[27,29,30,32] The small intensity arises because, with the present selection of 50% filling as the background filling since much of the volume that supports layer modes is excluded from the net $S(Q,\omega)$. Specifically, when $^4$He fills porous media at SVP, the $^4$He is deposited first on the media walls. Approximately the first 1.5 layers (to the extent one can speak of layers) are deposited as amorphous solid layers on the walls. Assuming 1.5 layers and cylindrical pores, the amorphous solid layers occupy 20 % of the pore volume. Next, liquid layers are deposited on the solid layers and filling proceeds from the walls inward. Measurements[19,40,54] of the filling dependence of $S(Q,\omega)$ show that the layer modes propagate in the first 2-5 liquid layers deposited on the amorphous solid. Assuming that the subsequent 2-5 liquid layers support layer modes, about the next 50% of the pore volume supports layer modes. However, only about 1/2 of this volume is included in the net $S(Q,\omega)$. It is for this reason that the layer mode intensity is weak relative to that of the roton in Fig. 2. The P-R mode propagates in both the initial liquid layers and in the interior of the pores (the remaining 25-30 % of the pore volume). The layer mode energy (mode center at 0.60 meV) and width (FWHM = 0.20 meV) at $Q = 1.92$ Å$^{-1}$ agrees with previous values[32] observed in MCM-41.

It would be interesting to determine the temperature dependence of $S(Q,\omega)$ in porous media at $Q$ values beyond the roton in future measurements. In bulk liquid $^4$He, the first two terms in Eq.(4) are observed in the superfluid phase only. In normal liquid helium, only $S_R(Q,\omega)$ is observed. There is transfer of intensity from the first two terms of $S(Q,\omega)$ to $S_R(Q,\omega)$ as temperature is increased into normal phase. At higher $\omega$, $S_R(Q,\omega)$ depends little on temperature. The loss of intensity in $S_1(Q,\omega)$ with increasing temperature tracks the condensate fraction well.[53] The intensity in $S_1(Q,\omega)$ goes to zero at $T_\lambda$ where $n_0$ goes to zero (i.e. $T_{BEC} = T_\lambda$ in bulk liquid helium). In porous media, we have found that P-R modes are observed at temperatures above $T_c$. That is, $T_{BEC} \geq T_c$ in porous media. It would be interesting to see whether the temperature dependence at $2 < Q < 3.6$ Å in porous media is the same as in the bulk but with a temperature $T_{BEC}$ that is greater than $T_c$.

## IV. CONCLUSION

The phonon-roton mode of liquid $^4$He confined in MCM-41 (47 Å diameter) and in the bulk have been measured under the same conditions at large wave vectors. The mode energy in fully filled MCM-41 and in the bulk liquid at SVP and at $T = 0.5$ K is the same within precision ($\pm 0.01$ meV) from $Q = 0.8$ Å$^{-1}$ out to the end point $Q = 3.6$ Å$^{-1}$. The mode width in MCM-41, as in bulk $^4$He, is too small to be observed at all $Q$ values (FWHM $\leq 0.02$ meV). The $Q$ dependence of the intensity, $Z_Q$, in the P-R mode is also the same and the present measurements provide new accurate values of $Z_Q$ in bulk in the range $2.3 \leq Q \leq 3.6$ Å$^{-1}$. The measurements show that the P-R mode energy in MCM-41 at higher wave vectors $Q$ remains sharply defined up to energy $2\Delta$ but the energy does not exceed $2\Delta$, where $\Delta$ is the roton energy. This shows that the P-R mode does not decay to layer modes since mode broadening would begin at a lower energy, e.g. $2\Delta_L$, and the mode energy would not reach $2\Delta$.

## V. ACKNOWLEDGEMENTS

It is a pleasure to acknowledge the support of the ISIS Facility, Rutherford Appleton Laboratory, UK and Richard Down for valuable technical assistance at ISIS. This work was supported by the DOE, Office of Basic Energy Sciences under contract No. ER46680.